\documentclass[letterpaper,12pt]{article}
\usepackage{amsmath}
\usepackage{amsfonts}
\usepackage{amssymb}
\usepackage{graphicx}
\usepackage{physics}
\usepackage{latexsym}
\usepackage{epsfig}
\usepackage{pstricks}
\usepackage{stmaryrd}
\usepackage{rotating}
\usepackage[english]{babel}
\usepackage{setspace}
\usepackage[utf8]{inputenc}
\usepackage{natbib}
\usepackage[T1]{fontenc}
\usepackage{lmodern}
\usepackage{enumitem}
\usepackage{csquotes}
\usepackage{amsthm}
\usepackage{xcolor}
\usepackage[margin=1in]{geometry}
\usepackage{comment}
\usepackage{hyperref}
\hypersetup{
	colorlinks=true,
	linkcolor=black,
	citecolor=teal,
	urlcolor=cyan,
}
\usepackage{enumitem}
\begin{document}
%%%%%%%%%%%%%%%%%%%%%%%%%%%%%%%%%%%%%%%%%%%%%%%%%%%%%%%%%%%%%%
%%%%%%%%%%%%%%%%%%%%%%%%%%%%%%%%%%%%%%%%%%%%%%%%%%%%%%%%%%%%%% 	
\begin{center}	
	\begin{Large}
		\textbf{Demystifying Relativistic Quantum Collapse}\\
	\end{Large}
\end{center}
	
\begin{center}
	\begin{large}
		R. Muciño, E. Okon, D. Sudarsky and M. Wiedemann\\
	\end{large}
	\textit{Universidad Nacional Aut\'onoma de M\'exico, Mexico City, Mexico.}\\[1cm]
\end{center}
%%%%%%%%%%%%%%%%%%%%%%%%%%%%%%%%%%%%%%%%%%%%%%%%%%%%%%%%%%%%%%
%%%%%%%%%%%%%%%%%%%%%%%%%%%%%%%%%%%%%%%%%%%%%%%%%%%%%%%%%%%%%%

Non-relativistic objective collapse theories have been remarkably successful in addressing the conceptual problems of standard quantum mechanics. Despite substantial efforts, the project of extending them to the relativistic domain remains burdened by significant conceptual objections and technical challenges, often taken to cast doubt on the viability of the program as a whole. On the conceptual side, relativistic collapse theories have been claimed to face challenges involving tension between instantaneous collapse and relativity, frame-dependence of property values and probabilities, the possibility of superluminal signaling and the failure of narratability. On the technical side, persistent infinities, difficulties in constructing fully covariant frameworks and the apparent need for non-standard fields have hindered the development of workable models. In this paper, we offer a systematic rebuttal of the conceptual objections and provide a structured account of the remaining technical challenges. We conclude that relativistic collapse theories do provide a promising route toward a fully relativistic quantum framework that overcomes the conceptual limitations of standard quantum theory.

%%%%%%%%%%%%%%%%%%%%%%%%%%%%%%%%%%%%%%%%%%%%%%%%%%%%%%%%%%%%%%
%%%%%%%%%%%%%%%%%%%%%%%%%%%%%%%%%%%%%%%%%%%%%%%%%%%%%%%%%%%%%%
\onehalfspacing
%\tableofcontents
%%%%%%%%%%%%%%%%%%%%%%%%%%%%%%%%%%%%%%%%%%%%%%%%%%%%%%%%%%%%%%
%%%%%%%%%%%%%%%%%%%%%%%%%%%%%%%%%%%%%%%%%%%%%%%%%%%%%%%%%%%%%%
\section{Introduction}
%%%%%%%%%%%%%%%%%%%%%%%%%%%%%%%%%%%%%%%%%%%%%%%%%%%%%%%%%%%%%%
%%%%%%%%%%%%%%%%%%%%%%%%%%%%%%%%%%%%%%%%%%%%%%%%%%%%%%%%%%%%%%

Non-relativistic objective collapse theories have been highly successful in addressing the conceptual problems of standard quantum mechanics. They introduce stochastic and nonlinear deviations from the Schrödinger equation, with the aim of providing a unified account of both the quantum behavior of microscopic systems and the absence of macroscopic superpositions, without appealing to observers or measurements.

The relativistic case, however, remains much more contentious. Despite substantial efforts to develop collapse models in the relativistic domain, there is still a widespread perception that objective collapse theories are fundamentally incompatible with relativity. This perception is based on a combination of conceptual and technical worries. Together, these worries have often been taken to suggest that the success of collapse theories in the non-relativistic setting cannot be extended to a fully relativistic quantum framework.

On the conceptual side, the worry is that a real collapse is difficult to reconcile with the relativistic structure of spacetime. If collapse is instantaneous, it seems to require a privileged notion of simultaneity; moreover, it appears to make property acquisition, stochasticity, and even the quantum state itself depend on the choice of foliation. Furthermore, the overt nonlocality of collapse has been thought to allow for superluminal signaling, potentially leading to causal paradoxes. Finally, because different foliations may yield different state histories, relativistic collapse appears to endanger \emph{narratability}: the possibility of giving a single, frame-independent account of what physically happens.

On the technical side, the construction of workable relativistic collapse models has also faced serious obstacles. Persistent infinities, difficulties in constructing fully covariant frameworks, and concerns about superluminality have all hindered progress. These issues are not merely matters of presentation or interpretation; they concern the formal and physical viability of the models themselves. Any satisfactory relativistic collapse theory must therefore meet demanding technical constraints, in addition to addressing the conceptual objections.

The aim of this paper is to reassess this pessimistic picture. We argue, first, that the main conceptual objections do not establish a fundamental incompatibility between objective collapse and relativity. Although the concerns involving preferred foliations, properties, probability, signaling or narratability raise important questions, they do not by themselves show that relativistic collapse theories are incoherent. Rather, they identify issues that must be handled carefully in formulating and interpreting such theories.

Second, we provide a structured account of the remaining technical hurdles. Persistent infinities, covariance requirements and signaling constraints pose substantial challenges for model construction, but they do not amount to reasons for abandoning the collapse program. A clearer assessment requires separating these difficulties, identifying their sources, and examining possible strategies for addressing them.

In what follows, we offer a systematic rebuttal of the conceptual objections and a structured account of the technical challenges that remain. Our conclusion is that relativistic collapse theories provide one of the most promising routes toward a fully relativistic quantum framework that overcomes the conceptual limitations of standard quantum theory.

The paper proceeds as follows. We begin, in section~\ref{NROCT}, with an overview of non-relativistic objective collapse theories. In section~\ref{CO}, we present the main conceptual objections that have been raised over the years regarding the construction of relativistic collapse models, and in section~\ref{ACI} we offer responses to those concerns. Next, in section~\ref{TC}, we summarize the technical challenges involved in constructing relativistic collapse models, and in section~\ref{SATI} we assess them. Finally, section~\ref{Con} contains our conclusions.

%%%%%%%%%%%%%%%%%%%%%%%%%%%%%%%%%%%%%%%%%%%%%%%%%%%%%%%%%%%%%%
%%%%%%%%%%%%%%%%%%%%%%%%%%%%%%%%%%%%%%%%%%%%%%%%%%%%%%%%%%%%%%
\section{Non-relativistic objective collapse theories}
\label{NROCT}
%%%%%%%%%%%%%%%%%%%%%%%%%%%%%%%%%%%%%%%%%%%%%%%%%%%%%%%%%%%%%%
%%%%%%%%%%%%%%%%%%%%%%%%%%%%%%%%%%%%%%%%%%%%%%%%%%%%%%%%%%%%%%

Non-relativistic objective collapse theories have been highly successful in addressing the conceptual problems of standard quantum mechanics, \cite{bassi2003dynamical,Bassi2013Models}. The central strategy is to modify the standard quantum dynamics by introducing stochastic and nonlinear deviations from the evolution dictated by the Schrödinger equation. These deviations are not tied to measurements or acts of observation; rather, they are taken to be part of the fundamental physical dynamics. The aim is to explain, within a single, perfectly precise framework, how microscopic systems can display characteristically quantum behavior while macroscopic objects always appear to occupy well-defined positions.\footnote{Given that collapse theories are designed to operate in settings where no observer or measuring device is involved, they are especially attractive for applications to non-standard scenarios, such as the early universe or the interior of black holes (see \cite{Benefits}).} 

To be empirically viable, however, such theories must satisfy certain constraints. First, the statistics of measurement outcomes must reproduce, or at least closely approximate, those given by the Born rule. Second, the energy increase produced by the non-unitary dynamics must remain within experimental bounds. Finally, collapse theories must be supplemented with local beables: a clear account of the physical entities that, according to the theory, exist in spacetime. This is needed because the wave function is defined on configuration space and, if included as part of the ontology, would have to be regarded as a nonlocal beable. Without a local ontology, the theory would specify a precise dynamics for the quantum state, but would not yet say clearly how that represents a physical state of affairs in spacetime.

Summing up, successful collapse models must satisfy the following requirements:
\begin{enumerate}
\item \textbf{Collapse dynamics}: the dynamics of the quantum state must include both unitary evolution and stochastic, nonlinear terms.
\item \textbf{Microscopic quantum behavior}: predictions for microscopic systems must reproduce known quantum behavior.
\item \textbf{Macroscopic definiteness}: macroscopic objects must always appear to occupy well-defined positions.
\item \textbf{Born-rule statistics}: measurement statistics must reproduce, or at least closely approximate, the Born rule.
\item \textbf{Bounded energy increase}: the energy increase produced by the non-unitary evolution must remain within experimental bounds.
\item \textbf{Local beables}: the theory must introduce local beables, specifying what physical entities exist in spacetime according to the theory.
\end{enumerate}

GRW, the simplest non-relativistic collapse model \cite{GRW}, is formulated for a system of $N$ distinguishable particles. Its central postulate is that each particle undergoes, with mean rate $\lambda$, sudden and spontaneous localization processes. If the $i$-th particle is localized around the point $\mathbf{x}$, the state vector changes according to
\begin{equation}
\ket{\psi}
\longrightarrow
\frac{\hat{L}_{i}(\mathbf{x})\ket{\psi}}
{\|\hat{L}_{i}(\mathbf{x})\ket{\psi}\|},
\end{equation}
where the operator $\hat{L}_{i}(\mathbf{x})$ is given by
\begin{equation}
\hat{L}_{i}(\mathbf{x})
=
\frac{1}{\left(2 \pi r^2 \right)^{3/4}}\
\exp\left[-\frac{\left(\hat{\mathbf{q}}_{i}-\mathbf{x}\right)^2}{4 r^2} \right],
\end{equation}
with $\hat{\mathbf{q}}_{i}$ the position operator of the $i$-th particle and $r$ a new fundamental constant that sets the characteristic localization length. Thus, a collapse does not localize the particle into an exact point, but rather into a finite spatial region centered around $\mathbf{x}$, thereby allowing the associated increase in energy to be kept under control.

The point $\mathbf{x}$ around which the localization of the $i$-th particle occurs is selected randomly, through a probability density
\begin{equation}
P_i(\mathbf{x}) = \bra{\psi} \hat{L}_{i}(\mathbf{x})^2 \ket{\psi}.
\end{equation}

This rule ensures that regions in which the wave function has greater weight are more likely to be selected as collapse centers. In this sense, the GRW collapse probabilities are closely related to the Born rule, although they are now incorporated into the fundamental stochastic dynamics rather than introduced as a separate measurement postulate.

Finally, between two successive spontaneous localization processes, the system evolves according to the usual Schrödinger equation. The complete dynamics therefore consists of ordinary unitary evolution interrupted at random times by spontaneous localization events. The collapse rate $\lambda$ is chosen to be very small for each individual particle, so that microscopic systems are only rarely affected. However, for systems with many entangled constituents, the probability that at least one particle undergoes a collapse is very large, producing the amplification mechanism needed to suppress macroscopic superpositions.

The Continuous Spontaneous Localization model, or CSL \cite{CSL}, replaces the discontinuous jumps of GRW with a continuous stochastic evolution of the quantum state. Instead of postulating that the wave function undergoes sudden localization events at random times, CSL modifies the Schrödinger equation by adding a continuous noise term that gradually drives the state toward localized configurations. In more detail, the state evolves according to
\begin{equation}
\frac{d}{dt}\ket{\psi(t)}_w 
=
\left[
-i\hat{H}
- \frac{1}{4 \lambda}
\int d \mathbf{x} \,
\left(w(\mathbf{x},t)-2 \lambda \hat{A}(\mathbf{x})\right)^2
\right]\ket{\psi(t)}_w ,
\end{equation}
where $\lambda$ controls the strength of the collapse dynamics, while $w(x,t)$ is a white-noise field. The operator $\hat{A}(x)$ is a smeared number-density operator defined as
\begin{equation}
\label{ACSL}
\hat{A}(\mathbf{x})
=
\int d\mathbf{z} \,
\hat{N}(\mathbf{z})
g_a(\mathbf{x}-\mathbf{z}),
\end{equation}
where $\hat{N}(\mathbf{z})$ is the particle number-density operator and $g_a(\mathbf{x}-\mathbf{z})$ a smearing function with $a$ its characteristic length.

Finally, the probability density assigned to a particular noise realization is not given by a fixed external measure alone. Rather, it is weighted by the norm of the corresponding state
\begin{equation}
P(w) Dw = {}_w\!\braket{\psi(t)}_w Dw,
\end{equation}
with
\begin{equation}
Dw = \prod_{i,k} \frac{dw_i(t_k)}{\sqrt{2 \pi \lambda/\Delta t}}.
\end{equation}
This weighting plays the role of selecting, with the appropriate, Born-like probabilities, those noise histories that drive the state toward one of the possible localized outcomes. Thus, CSL preserves the main physical aim of GRW---the dynamical suppression of macroscopic superpositions---while replacing abrupt collapses with a continuous stochastic process.

The preceding discussion explains how these models address the measurement problem at the level of the dynamics, in the sense that their evolution is precise and does not appeal to vague notions such as measurements or observers. However, as mentioned above, a satisfactory theory must also provide a clear physical interpretation. Without such an interpretation, the formalism would remain merely mathematical: it would specify how the quantum state evolves, but not how it represents a state of affairs in physical space.

In other words, since the wave function is defined on configuration space, it is not straightforward to interpret it directly as a physical field. For this reason, it has been proposed that collapse theories be supplemented with a primitive ontology: an account of the physical entities, or ``stuff,'' that the theory takes to exist in spacetime. Two main proposals have been developed in this regard. According to the flash ontology, the fundamental material entities are discrete spacetime events associated with the collapses of the GRW dynamics. On this view, matter is not represented by a continuous distribution, but by a pattern of localized events, or \emph{flashes}, occurring in spacetime. Macroscopic objects are then understood as stable patterns in the distribution of such flashes.

According to the mass-density ontology, by contrast, matter is represented by a continuous field in ordinary space. The mass density at a point $\mathbf{x}$ and time $t$ is defined by
\begin{equation}
m(x,t)
=
\bra{\psi(t)}\hat{M}(\mathbf{x})\ket{\psi(t)},
\end{equation}
where $\hat{M}(\mathbf{x})$ is the mass-density operator. Under this interpretation, the quantum state determines how matter is distributed in spacetime. The collapse dynamics then explains why, for macroscopic systems, this distribution becomes effectively localized rather than remaining spread over macroscopically distinct alternatives.

%%%%%%%%%%%%%%%%%%%%%%%%%%%%%%%%%%%%%%%%%%%%%%%%%%%%%%%%%%%%%%
%%%%%%%%%%%%%%%%%%%%%%%%%%%%%%%%%%%%%%%%%%%%%%%%%%%%%%%%%%%%%%
\section{Conceptual objections}
\label{CO}
%%%%%%%%%%%%%%%%%%%%%%%%%%%%%%%%%%%%%%%%%%%%%%%%%%%%%%%%%%%%%%
%%%%%%%%%%%%%%%%%%%%%%%%%%%%%%%%%%%%%%%%%%%%%%%%%%%%%%%%%%%%%%

The objective collapse models discussed above are non-relativistic. The relativistic case, by contrast, remains far more controversial. Despite substantial efforts to develop collapse models in the relativistic domain, there remains a widespread perception that objective collapse theories are fundamentally incompatible with relativity. The challenges arise from the apparent difficulty of understanding collapse as a real physical process within a relativistic spacetime. If collapse is objective, questions immediately arise about where and when it occurs and how its effects propagate. Over the years, many objections have been raised against the possibility of constructing a relativistic objective collapse model. In what follows, we organize and summarize these objections. 

%%%%%%%%%%%%%%%%%%%%%%%%%%%%%%%%%%%%%%%%%%%%%%%%%%%%%%%%%%%%%%
\subsection{Instantaneous collapses vs. relativity}
%%%%%%%%%%%%%%%%%%%%%%%%%%%%%%%%%%%%%%%%%%%%%%%%%%%%%%%%%%%%%%

A central worry is that, if collapse is understood as an instantaneous physical process, then it must occur along some spacelike hypersurface. This would seem to introduce a privileged foliation of spacetime, in tension with the relativistic absence of an absolute simultaneity relation. For instance, in an entangled two-particle system, the observation of one particle seems to induce a sudden change in the state of its distant partner. In a non-relativistic setting, this is usually described as an instantaneous collapse of the wave function. But in Minkowski spacetime there is no absolute fact about simultaneity between spacelike separated events. Thus, it is argued that, either the theory must identify a preferred foliation with respect to which collapse really occurs, or it must give up the idea that collapse is an instantaneous physical process.

%%%%%%%%%%%%%%%%%%%%%%%%%%%%%%%%%%%%%%%%%%%%%%%%%%%%%%%%%%%%%%
\subsection{Frame-dependence of property values}
%%%%%%%%%%%%%%%%%%%%%%%%%%%%%%%%%%%%%%%%%%%%%%%%%%%%%%%%%%%%%%

A further objection is that relativistic collapse seems to make the attribution of physical properties depend on the choice of frame. Consider two spin-1/2 particles in a singlet measured at spacelike separation by Alice and Bob. In a frame in which Alice measures first, her measurement collapses the state and Bob's particle is thereby assigned a definite spin before Bob measures it. In a frame in which Bob measures first, the reverse is true: Bob's measurement collapses the singlet state and Alice's particle acquires a definite spin state before Alice measures it. The disagreement is not merely about the spatial or temporal description of the events, as relativity allows, but about apparently non-spatiotemporal facts: whether a given particle has a definite spin at a given event. The worry is that it is hard to make sense of a particle at a given spacetime location having one spin status relative to one hypersurface and another relative to a different hypersurface. If spin possession is supposed to be a genuine physical fact, it seems that it should not depend on how spacetime is sliced. Thus, relativistic collapse appears to threaten a consistent, frame-independent account of property acquisition.

%%%%%%%%%%%%%%%%%%%%%%%%%%%%%%%%%%%%%%%%%%%%%%%%%%%%%%%%%%%%%%
\subsection{Frame-dependence of probability}
%%%%%%%%%%%%%%%%%%%%%%%%%%%%%%%%%%%%%%%%%%%%%%%%%%%%%%%%%%%%%%

A related worry concerns the modal status of detection events. As we just saw, in a frame in which Alice measures first, her measurement collapses the singlet state, so Bob's later result may be regarded as determined by the spin state already assigned to his particle. In a different frame, however, Bob measures first, so it is Alice's later result that is determined by the prior collapse. Thus, different frames appear to disagree about which detection event was genuinely stochastic and which merely revealed an already definite spin value. The objection is that whether an event is chancy or determined should not depend on the choice of reference frame; yet relativistic collapse seems to make precisely that distinction hypersurface-dependent.

An associated way of presenting this objection is as follows. Collapse models are fundamentally indeterministic: the occurrence, location, or outcome of a collapse is not fixed in advance by the prior quantum state. This may appear to sit uneasily with the block-universe picture associated with relativity, according to which spacetime is given as a four-dimensional whole rather than generated through an objective temporal becoming. If collapse involves genuine stochastic events, one may wonder how such events fit into a relativistic spacetime picture in which there is no preferred global present.

%%%%%%%%%%%%%%%%%%%%%%%%%%%%%%%%%%%%%%%%%%%%%%%%%%%%%%%%%%%%%%
\subsection{The quantum state cannot be objective}
%%%%%%%%%%%%%%%%%%%%%%%%%%%%%%%%%%%%%%%%%%%%%%%%%%%%%%%%%%%%%%

The next concern can be formulated as follows. The frame-dependence of collapse seems to undermine any interpretation of the quantum state as a real, observer-independent physical state or object. If different foliations assign different collapsed states to the same physical situation, then the quantum state appears to depend not only on what exists in the world, but also on the hypersurface relative to which the situation is described. This makes it difficult to regard the quantum state itself as an objective element of reality, rather than as an epistemic device representing perspectival information.

%%%%%%%%%%%%%%%%%%%%%%%%%%%%%%%%%%%%%%%%%%%%%%%%%%%%%%%%%%%%%%
\subsection{Signaling}
%%%%%%%%%%%%%%%%%%%%%%%%%%%%%%%%%%%%%%%%%%%%%%%%%%%%%%%%%%%%%%

Signaling raises a further concern. The collapse process is overtly nonlocal: changes associated with one part of an entangled system can affect the state assigned to distant parts of that system. This might seem to open the door to superluminal signaling, together with the causal paradoxes usually associated with faster-than-light communication. Even if nonlocal collapse is intended to reproduce the empirical predictions of quantum mechanics, the worry is that making collapse a real dynamical process could turn harmless quantum nonlocality into a controllable channel for transmitting information outside the light cone.

%%%%%%%%%%%%%%%%%%%%%%%%%%%%%%%%%%%%%%%%%%%%%%%%%%%%%%%%%%%%%%
\subsection{Narratability}
%%%%%%%%%%%%%%%%%%%%%%%%%%%%%%%%%%%%%%%%%%%%%%%%%%%%%%%%%%%%%%

For a classical system, specifying the history of the state along one foliation uniquely determines the corresponding history along any other foliation, independently of the system's dynamics. David Albert has called this phenomenon \emph{narratability}, \cite{Albert2015}. Quantum state evolution does not generally have this feature: there are cases in which a state history along one foliation is compatible with several distinct state histories along another. Then the theory seems unable to provide a single, frame-independent account of what physically happens: real state reduction appears difficult to reconcile with a coherent relativistic history of the quantum state.

%%%%%%%%%%%%%%%%%%%%%%%%%%%%%%%%%%%%%%%%%%%%%%%%%%%%%%%%%%%%%%
\subsection{The Esfeld-Gisin argument}
%%%%%%%%%%%%%%%%%%%%%%%%%%%%%%%%%%%%%%%%%%%%%%%%%%%%%%%%%%%%%%

We close this list of objections with an argument by Esfeld and Gisin in \cite{EsfeldGisin2014GRWFlash}, that purports to establish the impossibility of a relativistic collapse model. Although the argument is presented by the authors specifically against a GRW model with a flash ontology, it does not appear to rely on any distinctive features of that theory, and may therefore be read as a more general argument against relativistic collapse.

In an EPR setting, let $x$ denote Alice's measurement setting and $a$ her outcome, and let $y$ denote Bob's measurement setting and $b$ his outcome. Consider first a reference frame in which Alice's measurement occurs before Bob's. In such a frame, Alice's outcome may be represented as
\begin{equation}
a = F_{AB}(x,\lambda),
\end{equation}
where $\lambda$ denotes the relevant background variables. Since Bob is second in this temporal ordering, his outcome may depend not only on his own setting $y$, but also on Alice's setting and outcome:
\begin{equation}
b = S_{AB}(x,y,a,\lambda).
\end{equation}
Here $F_{AB}$ denotes the outcome function for the first measurement in the $A$-then-$B$ ordering, while $S_{AB}$ denotes the outcome function for the second measurement in that ordering.

Now consider a different reference frame, in which Bob's measurement occurs before Alice's. By the same reasoning, Bob's outcome may be written as
\begin{equation}
b = F_{BA}(y,\lambda),
\end{equation}
while Alice's outcome, now second in the temporal ordering, may be written as
\begin{equation}
a = S_{BA}(y,x,b,\lambda).
\end{equation}
For the actual outcomes to be Lorentz invariant, they must not depend on which reference frame is used to describe their temporal ordering. Thus, the following equalities would have to hold:
\begin{equation}
a = F_{AB}(x,\lambda) = S_{BA}(y,x,b,\lambda),
\end{equation}
and
\begin{equation}
b = S_{AB}(x,y,a,\lambda) = F_{BA}(y,\lambda),
\end{equation}
for all possible measurement settings $x$ and $y$.

The problem is that no functions $F_{AB}$, $F_{BA}$, $S_{AB}$, and $S_{BA}$ satisfying these conditions can reproduce the quantum correlations. In particular, the second equality implies that $S_{AB}$ must be independent of Alice's setting $x$ and outcome $a$. But then $\lambda$ would function as a local hidden variable in Bell's sense. Since local hidden variables cannot reproduce correlations that violate Bell inequalities, the required functions cannot exist.

%%%%%%%%%%%%%%%%%%%%%%%%%%%%%%%%%%%%%%%%%%%%%%%%%%%%%%%%%%%%%%
\subsection{Summary of objections}
%%%%%%%%%%%%%%%%%%%%%%%%%%%%%%%%%%%%%%%%%%%%%%%%%%%%%%%%%%%%%%

All these objections can be summarized as follows:
\begin{description}%[labelsep=.5cm]
\item[A.] \textbf{Instantaneous collapses vs. relativity}: if collapse is a real instantaneous physical process, then it seems to require a privileged notion of simultaneity, in tension with relativity.
\item[B.] \textbf{Frame-dependence of property values}: relativistic collapse seems to make property acquisition frame-dependent, so that different frames disagree about whether a system possesses a definite property.
\item[C.] \textbf{Frame-dependence of probability}: relativistic collapse seems to make it frame-dependent whether a detection event is genuinely stochastic or determined by a prior collapse.
\item[D.] \textbf{The quantum state cannot be objective}: the frame-dependence of a relativistic collapse seems to undermine any interpretation of the quantum state as a real, observer-independent physical state.
\item[E.] \textbf{Signaling}: the overt nonlocality of collapse might allow superluminal signaling, together with the associated causal paradoxes.
\item[F.] \textbf{Narratability}: relativistic collapse appears to threaten \emph{narratability}, making it difficult to reconcile with a single, frame-independent account of what physically happens.
\item[G.] \textbf{The Esfeld-Gisin argument}.
\end{description}

%%%%%%%%%%%%%%%%%%%%%%%%%%%%%%%%%%%%%%%%%%%%%%%%%%%%%%%%%%%%%%
%%%%%%%%%%%%%%%%%%%%%%%%%%%%%%%%%%%%%%%%%%%%%%%%%%%%%%%%%%%%%%
\section{Addressing the conceptual objections}
\label{ACI}
%%%%%%%%%%%%%%%%%%%%%%%%%%%%%%%%%%%%%%%%%%%%%%%%%%%%%%%%%%%%%%
%%%%%%%%%%%%%%%%%%%%%%%%%%%%%%%%%%%%%%%%%%%%%%%%%%%%%%%%%%%%%%

The responses to the conceptual objections developed below are inspired by, and in some cases directly based on, earlier work by Aharonov, Albert, Pearle, Ghirardi, Myrvold, Bedingham, and others (see, e.g., \cite{AAI,AAII,Pearle1990,GGP1990,GP1991,Albert2000,myrvold2002peaceful,Myrvold,bassi2003dynamical, bedingham2011relativistic,bedingham2014matter,Myrvold2021}). Our aim is not to present these responses as new, but to organize them into a unified framework and to show how they bear on the alleged incompatibility between objective collapse and relativity.

Before turning to the objections, though, it will be useful to state explicitly what we take a successful relativistic objective collapse model to require. These requirements come on top of the general constraints on collapse theories listed above: reproducing quantum behavior at the microscopic level, suppressing macroscopic superpositions, approximating the Born rule, keeping energy increase within experimental bounds, and providing a clear ontology of local beables. The point is to identify the additional constraints that arise specifically when the collapse program is extended to a relativistic spacetime.

First, the theory must be formulated in terms appropriate to relativistic physics. In a relativistic setting, there is no distinguished global time parameter with respect to which the quantum state simply evolves. A natural replacement is to assign quantum states to Cauchy hypersurfaces, so that the state represents the physical situation relative to a spacelike slice of spacetime. This allows one to describe the evolution of the quantum state without presupposing a preferred temporal foliation.

Second, the dynamics of the state and the construction of the local beables must be covariant. If the theory is to be genuinely relativistic, neither the collapse dynamics nor the ontology should depend on extra spacetime structure, such as a preferred foliation. The metric may of course enter the formulation, since it is part of the relativistic spacetime structure. But the theory should not rely on additional absolute structures that would reintroduce, by hand, something like a preferred simultaneity relation.

Third, the state assignments associated with different Cauchy hypersurfaces must be compatible (see \cite[sec. 4.2]{Myrvold2019}). In a relativistic collapse theory, a compact spacelike region $\alpha$ may be contained in many different Cauchy hypersurfaces. Since collapses occurring at spacelike separation from $\alpha$ can lie to the past of some of these hypersurfaces but not others, the reduced states assigned to $\alpha$ from those hypersurfaces need not coincide. This disagreement is not by itself problematic: such reduced states are extrinsic, hypersurface-relative state assignments, not rival descriptions of the \emph{intrinsic} physical state of $\alpha$. What would be problematic is if some of them were in outright conflict, for example by having orthogonal supports\footnote{In the finite dimensional case, the support $S(\rho)$ of a density matrix $\rho$ is the subspace spanned by all its eigenvectors with non-zero eigenvalues \cite{Brun}. For the infinite dimensional version, see \cite[Appendix]{Myrvold2019}.} and thereby assigning probability one to incompatible outcomes of some possible local experiment. A natural consistency requirement is therefore that, for any compact spacelike region, the reduced states associated with all hypersurfaces containing that region have overlapping support . In that case, different hypersurface-relative state assignments may be different, but they do not yield incompatible accounts of what can occur in the region.

Finally, the theory should satisfy a no-signaling condition. The nonlocality of collapse is not by itself fatal: ordinary quantum theory already contains nonlocal correlations. What would be problematic, at least by standard lights, is if this nonlocality could be used to transmit controllable information outside the light cone. One could consider relaxing this requirement if it could be shown that the resulting theory avoids causal paradoxes despite allowing some form of superluminal influence or signaling. In the absence of such a demonstration, however, no-signaling remains a natural constraint on any viable relativistic collapse model.

Summing up, in addition to the general requirements listed above, a successful relativistic objective collapse model should satisfy the following:
\begin{enumerate}
\setcounter{enumi}{6}
\item \textbf{Relativistic quantum states}: a quantum state is assigned to every Cauchy hypersurface of a globally hyperbolic spacetime.
\item \textbf{Covariance}: the dynamics of the state and the construction of the local beables must not depend on any spacetime structure beyond the metric.
\item \textbf{Consistency}: for any compact spacelike region, the reduced states associated with all hypersurfaces containing that region must be compatible, in the sense that the intersection of the supports of the outcome probabilities (extracted from the reduced states) is non-empty.

\item \textbf{No-signaling}: local observations must not allow one to determine whether operations have been carried out at spacelike separation.
\end{enumerate}

Below, we will discuss concrete attempts at constructing models satisfying these requirements. Before doing so, however, we explain how any model that satisfies them is able to address the conceptual objections raised above.

We begin with the alleged tension between instantaneous collapse and relativity. Suppose that we have a model in which every Cauchy hypersurface is assigned a quantum state and, moreover, that the procedure determining this assignment---that is, the dynamics of the state---is covariant. In that case, the collapse dynamics no longer requires a preferred foliation, and the central source of the alleged conflict with relativity is removed. The question, of course, is whether such a construction can actually be achieved. To see that it can, consider the strategy proposed by Aharonov and Albert in \cite{AAII}, which is available for models with discrete collapses. The idea is not to say that collapses affect states when the leaves of a preferred foliation sweep across them, but rather, in a Tomonaga-Schwinger type of framework, that states are affected when an \emph{arbitrary} spacelike hypersurface is continuously deformed across the collapse event. In this way, a manifestly Lorentz-invariant version of the collapse postulate can be formulated: collapse occurs as any spacelike hypersurface is advanced through the relevant collapse event.

This removes the immediate tension between instantaneous collapse and relativity. However, it seems to leave wide open the door for problems arising from a frame-dependence of property values, i.e., objection B. That is, with this collapse recipe, it will be the case that the states associated with all the different hypersurfaces going through some event will, in general, be different. For instance, on a hypersurface that passes just before Bob's measurement and lies to the past of Alice's, the state is a singlet; but on another hypersurface that also passes just before Bob's measurement while lying to the future of Alice's, the state is separable. The worry is that whether a particle possesses a definite spin seems like a genuine physical fact, not something that should depend on the choice of hypersurface.

The response is that this objection conflates two different things: the quantum state assigned to a hypersurface and the \emph{local beables} associated with a bounded spacetime region. The quantum state is not a local beable that acquires values at spacetime points; it is a non-local beable assigned to an entire Cauchy hypersurface. As such, there is no reason to expect two hypersurfaces that cross the same local region to carry the same quantum state. They are not two competing values of a local field at the crossing point.

If one wants to ask what physical properties are instantiated in the overlap region, one must instead look at the local beables. These are precisely introduced to say what, according to the theory, exists in ordinary spacetime. Thus, the relevant question is not whether the quantum states assigned to all hypersurfaces through a region are identical, but whether the theory gives a consistent assignment of local beables in that region. If the local beables are constructed covariantly, as required above, then their assignment cannot depend on an arbitrary choice of foliation. In that case, no inconsistency in property attribution arises merely from the fact that different hypersurfaces carry different quantum states.

This answers objection B. Relativistic collapse need not make local property possession frame-dependent, provided that physical properties are read from covariantly defined local beables rather than from the global quantum state assigned to a hypersurface. Still, the discussion points toward a further problem. Even if local property values are not read directly from the hypersurface-relative quantum state, probabilities are typically constructed from such states. Thus, while the frame-dependence of the quantum state need not generate inconsistent property assignments, it may still seem to threaten a consistent account of which events are stochastic and which are determined, or of the assignment of probabilities in general. That is the issue raised by objection C.

A similar response applies. In a relativistic setting, in which one must accept the block-universe picture, probabilities should not be understood as marking an objective process of becoming in which future events are progressively created. From the standpoint of the whole spacetime block, all events are already part of the complete four-dimensional history. Even if the dynamics is fundamentally indeterministic, the actual block contains a definite pattern of collapse events and outcomes. What remains meaningful is the probability of some event \emph{conditional} on a given physical state, or on the information encoded by that state.

Understood in this way, there is nothing surprising about probabilities changing from one hypersurface to another. Different hypersurfaces are associated with different quantum states because they contain, or have to their past, different information about the block. A state prior to Alice's measurement and a state posterior to Alice's measurement need not assign the same probabilities to Bob's outcome, because they represent different conditional standpoints. The difference is not a contradiction about what is objectively chancy; it is a difference in the information relative to which probabilities are being assigned.

Thus, the frame-dependence of probabilities is not by itself problematic. What must be required is that the various hypersurface-relative probability assignments be mutually consistent: they must not yield incompatible accounts of what can occur in the relevant spacetime region. This is precisely what the consistency requirement (item 9 above) is designed to secure. Once that condition is imposed, the fact that different hypersurfaces support different conditional probability assignments does not amount to an objection to relativistic collapse.

Objection D demanding an epistemic interpretation of the quantum state can be answered along the same lines. The fact that different hypersurfaces may be assigned different quantum states does not show that the quantum state is merely epistemic or observer-dependent. It shows only that, in a relativistic setting, the quantum state is a nonlocal beable assigned to an entire Cauchy hypersurface. Its objectivity consists in the fact that, given the physical history and the covariant collapse dynamics, the state associated with each hypersurface is fixed by the theory. Thus, the hypersurface-dependence of the quantum state is not a dependence on an observer's perspective or knowledge, but a structural feature of the relativistic formulation. What must be frame-independent are the rules assigning states to hypersurfaces and the local beables constructed from them. If those rules are covariant, then the quantum state can be treated as an objective nonlocal element of the theory without requiring it to be the same on all hypersurfaces passing through a given region.

Next, we consider objection E about signaling, which should be treated with care. First, the nonlocality involved in relativistic collapse is not an idiosyncratic feature of collapse theories. Bell's theorem shows that any framework capable of reproducing the quantum correlations must, in some sense, be nonlocal. The relevant question, therefore, is not whether objective collapse theories are nonlocal, but whether their nonlocality can be kept under control in such a way as to avoid unacceptable physical consequences. Since other nonlocal approaches to quantum theory can avoid operational signaling, there is no reason in principle to think that objective collapse theories could not do the same.

Moreover, even if one were to consider models that allow some form of superluminal signaling, it would not immediately follow that such models generate causal paradoxes, \cite{maudlin2011quantum}. The usual worry is that superluminal signaling, together with Lorentz invariance, permits closed causal loops. But this inference assumes that events can be altered along such loops in the way suggested by informal time-travel stories. In a relativistic block-universe picture, however, the events in spacetime form a single consistent history. Just as closed timelike curves need not generate paradoxes if what happens at a spacetime point is fixed once and for all, a covariant collapse model allowing superluminal signaling need not thereby be paradoxical: the relevant signals would have to be embedded in one consistent spacetime history.

Thus, no-signaling is a natural and conservative requirement for relativistic collapse models, but it is not obvious that it is conceptually mandatory. What is required is the absence of inconsistency. If a model satisfies covariance and yields a coherent assignment of events, states, and local beables throughout spacetime, then the presence of some controlled form of superluminal signaling would not by itself establish a contradiction. The burden would be to show that such signaling leads to inconsistent spacetime histories, not merely that it violates a familiar expectation about relativistic theories.

The narratability objection rests on a standard that need not be accepted as a general constraint on relativistic theories. In a theory whose ontology consists exclusively of local beables, as in classical relativistic field theories, a complete history along one foliation will indeed determine the corresponding history along any other foliation. But this is because the history along the first foliation already contains all local facts, which in this case are all the physical facts: specifying the state at every point of every leaf amounts to specifying what happens everywhere in spacetime. The history along another foliation can then be reconstructed simply by reorganizing the same collection of local facts.

The situation is different in a theory that includes nonlocal beables, such as the quantum state. Since the quantum state is assigned to entire hypersurfaces, and not pointwise to spacetime locations, the full state history along one foliation need not contain all the information required to reconstruct the state history along another. This is not a failure of relativistic coherence. It reflects the fact that the theory's ontology is not exhausted by local beables. Narratability, in the classical sense, is therefore not a generic requirement of relativistic physics; it is a special feature of theories whose complete ontology is local. A genuinely relativistic theory with nonlocal beables may fail to be narratable in Albert's sense without thereby being inconsistent, non-covariant, or physically unintelligible.

Finally, the Esfeld-Gisin argument can be answered by pointing out that the asymmetric dependence assumptions on which it relies are unwarranted. The argument assumes that in a frame in which, say, Alice's measurement is assigned the earlier time coordinate, Alice's outcome can depend only on her local setting and on $\lambda$, while Bob's outcome may also depend on Alice's setting and outcome. But this is not the correct lesson to draw from the relativistic treatment of simultaneity. For spacelike separated events, there simply is no invariant temporal order. The fact that Alice's measurement receives a lower time coordinate in one Lorentz frame does not license the conclusion that Alice's result is in any physical sense prior to Bob's, or that it cannot depend nonlocally on Bob's setting. Nor does the fact that Bob's measurement receives a higher time coordinate in that frame license the conclusion that Bob's outcome may depend on Alice's setting in a way unavailable to Alice's outcome.

Once this is recognized, the symmetry of the situation is restored. Regardless of the Lorentz frame used to describe the experiment, Alice's and Bob's measurements are spacelike separated. Neither is absolutely first, and neither is absolutely second. Therefore, a relativistic collapse theory need not represent the outcomes by functions in which the ``first'' result depends only on the local setting and $\lambda$, while the ``second'' result may depend on the distant setting and outcome. That distinction imports a frame-dependent temporal ordering into the dynamics, precisely what a covariant relativistic account should avoid.

Thus, the Esfeld-Gisin argument does not show that relativistic collapse is impossible. It shows only that one cannot combine Lorentz invariance with a frame-dependent picture in which whichever measurement is first in a chosen coordinate system is treated as dynamically independent of the spacelike separated measurement, while the second is allowed to depend on the first. But a genuinely relativistic treatment should not assign such asymmetric dynamical roles to spacelike separated events in the first place. All that can be said invariantly is that the two measurements are spacelike separated, and any admissible dependence structure must respect that symmetry.

%%%%%%%%%%%%%%%%%%%%%%%%%%%%%%%%%%%%%%%%%%%%%%%%%%%%%%%%%%%%%%
%%%%%%%%%%%%%%%%%%%%%%%%%%%%%%%%%%%%%%%%%%%%%%%%%%%%%%%%%%%%%%
\section{Technical challenges} 
\label{TC}
%%%%%%%%%%%%%%%%%%%%%%%%%%%%%%%%%%%%%%%%%%%%%%%%%%%%%%%%%%%%%%
%%%%%%%%%%%%%%%%%%%%%%%%%%%%%%%%%%%%%%%%%%%%%%%%%%%%%%%%%%%%%%

For some time now, it has been clear that relativistic collapse models can be constructed so as to satisfy the central relativistic requirements, as well as most of the general desiderata for objective-collapse theories. Early efforts in this direction focused mainly on extending CSL to quantum field theory. Pearle’s initial proposal \cite{Pearle1990}, later developed further by Ghirardi, Grassi, and Pearle \cite{GGP1990}, introduced CSL-type stochastic and nonlinear collapse terms within a relativistic quantum-field-theoretic framework. These models demonstrated that relativistic collapse dynamics could indeed be formulated with the desired structural features: covariance, consistency, no-signaling, as well as the adequate microscopic quantum behaviour, the recovery of the Born rule and the suppression of macroscopic superpositions. However, they also revealed a serious technical obstacle: the collapse dynamics explored produced an infinite increase of energy per unit time and unit volume in physical systems, making them physically unacceptable despite their conceptual and relativistic promise.

To illustrate generically how models of this kind work, let $\ket{\Psi_\Sigma}$ denote the quantum state of the matter fields associated with the spacelike hypersurface $\Sigma$. We work in a hybrid picture in which the unitary evolution, both free and interacting, is encoded in the Heisenberg evolution of the field operators, while the collapse dynamics is represented by the evolution of the hypersurface-dependent state.

In a CSL-type, continuous version, the change in the state, as the hypersurface is advanced toward the future along an arbitrary foliation of spacetime, is governed by a Tomonaga-Schwinger equation
\begin{equation}
\label{RCSL}
\frac{\delta \ket{\Psi_\Sigma}_w}{\delta \Sigma(x)}
=-\frac{1}{4 \lambda}
\left[w(x) - 2 \lambda \hat{T}(x) \right]^2 \ket{\Psi_\Sigma}_w,
\end{equation}
with $w(x)$ a Gaussian white noise with zero mean, ensemble average
\begin{equation}
\langle \! \langle w(x) w(x') \rangle \! \rangle = \lambda \delta (x-x')
\end{equation}
and probability density
\begin{equation}
P(w) Dw = {}_w\!\braket{\Psi_\Sigma}_w Dw .
\end{equation}
Covariance, understood as foliation-independence of the dynamics, is guaranteed by the condition 
\begin{equation}
\label{AA}
[\hat{T}(x),\hat{T}(y)] = 0,
\end{equation}
whenever $x$ and $y$ are spacelike separated. Moreover, by taking $\hat{T}(x)$ to be an appropriate relativistic generalization of the non-relativistic mass-density operator, i.e., some operator associated with the energy-momentum tensor at point $x$ (e.g., its trace),\footnote{For a procedure to deal with the renormalization of the quantum stress tensor fluctuations see \cite{PerezSudarsky}.} the model gradually drives the state toward configurations of well-defined energy-momentum, suppressing undesired macroscopic superpositions.

In a GRW-type, discrete-collapse version, one extends the unitary dynamics by stipulating that the state undergoes discrete collapses, associated with randomly selected spacetime points assumed to occur with constant probability per unit four-volume---note that such a distribution of collapse events is fully covariant. Collapses are implemented by specifying that, when a hypersurface $\Sigma$ passes through a collapse point $x$, the state transforms according to
\begin{equation}
\label{CD}
\ket{\Psi_\Sigma}
\longrightarrow
\frac{\hat{L}_x(Z_x)\ket{\Psi_\Sigma}}{\|\hat{L}_x(Z_x)\ket{\Psi_\Sigma}\|},
\end{equation}
with $\hat{L}_x(Z_x)$ an un-sharp projection into an eigenstate of $\hat{T}(x)$ with eigenvalue $Z_x$
\begin{equation}
\label{Lx}
\hat{L}_x(Z_x)
=
\frac{1}{(2\pi\sigma^2)^{1/4}}
\exp\left[
-\frac{(\hat{T}(x)-Z_x)^2}{4\sigma^2}
\right].
\end{equation}
The $Z_x$ are random variables, selected according to the probability distribution
\begin{equation}
\label{ProbZ}
P(Z_x) = \bra{\Psi_\Sigma}\hat{L}_x(Z_x)^2\ket{\Psi_\Sigma},
\end{equation}
which requires the $\hat{L}_x$ to satisfy the completeness condition
\begin{equation}
\label{Comp}
\int \, \hat{L}_x(z)^2 dz = I .
\end{equation}
Since collapses constitute quasi-projections onto approximate eigenstates of $\hat{T}(x)$, the cumulative effect of many collapses is to drive the system toward eigenstates of $\hat{T}(x)$. As a result, the model also displays the desired property of suppressing undesired macroscopic superpositions.

Since the $\hat{T}$ operators satisfy Eq.~\eqref{AA} for spacelike separated points $x$ and $y$, given initial and final hypersurfaces $\Sigma_i$ and $\Sigma_f$ with no point of $\Sigma_i$ to the future of $\Sigma_f$, the dynamics assigns a unique state to $\Sigma_f$, independently of the foliation connecting the two hypersurfaces. That is, each Cauchy hypersurface is assigned a state, and the assignment is fully covariant. Moreover, satisfaction of the microcausality-like Eq.~\eqref{AA} guarantees the absence of superluminal signaling, both in the discrete and continuous cases.

To complete the description of these frameworks, the associated local ontology must be specified. In this regard, following \citep{bedingham2011relativistic,bedingham2014matter}, one can introduce an energy-momentum density $\mathcal{T}_{ab}(x)$, defined by the renormalized expectation value of the energy-momentum tensor operator. Since, in these collapse theories, the expectation value at a point depends on the hypersurface chosen, the relevant hypersurface must be specified. The proposal is to calculate it in the state assigned to the boundary of the causal past of $x$, $\partial J^-(x)$:
\begin{equation}
\label{matterdensityeq}
\mathcal{T}_{ab}(x)
\equiv
\langle \psi | \hat{T}_{ab} | \psi \rangle^{\text{Ren}}_{\partial J^-(x)} ,
\end{equation}
As required, this prescription is fully covariant. Moreover, it seems both a natural generalization of the mass-density ontology and a particularly natural choice if the scheme is to be used in conjunction with a semiclassical treatment of gravitation (see \cite{Fully}).

The constructions presented above have many of the features required of a successful relativistic collapse framework listed above. They provide a relativistic collapse dynamics formulated on arbitrary spacelike hypersurfaces, the dynamics and the definition of the local ontology are covariant, the predictions are consistent and reproduce many of the successes of the non-relativistic models. However, a serious difficulty emerges when one considers the expected change in energy generated by the collapse dynamics: the expectation value of the energy changes by an infinite amount. The source of the divergence is the pointlike character of the $\hat{T}$ operators, which are too singular to be used directly in the Tomonaga-Schwinger Eq.~\eqref{RCSL} or in the construction of the $\hat{L}$ quasi-projectors of Eq.~\eqref{Lx}. As a result, the collapse dynamics produces arbitrarily sharp variations in the field configuration, and these discontinuities contribute an infinite amount to the energy.

The solution to this problem seems quite natural. As is already the case in the collapse operator for the non-relativistic CSL model (see Eq.~\eqref{ACSL}), one should replace the pointlike $\hat{T}$ by a smeared version, spread over some finite spacetime region by means of an appropriate smearing function. Such a modification would soften the discontinuities and thereby avoid the infinite energy increase. However, this move introduces two difficulties. First, it is not straightforward to define the smearing function in a genuinely Lorentz-covariant way, since many natural smearing prescriptions implicitly pick out a preferred frame or foliation. Second, even if a suitable covariant smearing function can be found, the modified collapse operators may no longer satisfy Eq.~\eqref{AA}. And if those equations fail, the evolution may threaten both covariance and no-signaling.

In sum, the technical challenge before us is to construct a framework that, simultaneously, satisfies \textbf{bounded energy increase}, \textbf{covariance}, and \textbf{no-signaling}.

%%%%%%%%%%%%%%%%%%%%%%%%%%%%%%%%%%%%%%%%%%%%%%%%%%%%%%%%%%%%%%
\subsection{Myrvold's no-go result}
%%%%%%%%%%%%%%%%%%%%%%%%%%%%%%%%%%%%%%%%%%%%%%%%%%%%%%%%%%%%%%

This tension between covariance, signaling and stability is formalized by Myrvold in \cite{myrvold2017}, with a no-go result alleging that, if all degrees of freedom of relativistic quantum field theory are \emph{standard}, vacuum stability together with the requirements imposed by relativistic causality entails deterministic evolution.

In more detail, it is argued for the following: a relativistic, Markovian, stochastic collapse theory in Minkowski spacetime cannot be built using only the ordinary degrees of freedom of standard relativistic quantum field theory. If such a theory is to work, it must introduce something like the nonstandard pointer fields used by Bedingham \cite{bedingham2011relativistic} and Pearle \cite{Pearle2015}.

The argument has three main parts. The first step is to introduce a general framework for Markovian stochastic quantum evolution. In a stochastic Tomonaga-Schwinger picture of the kind employed above, the collapse dynamics is represented by completely positive maps, or equivalently by a family of Kraus operators, as follows. For each pair of Cauchy surfaces $\Sigma$ and $\Sigma'$, with $\Sigma'$ nowhere to the past of $\Sigma$, let $\delta$ be the spacetime region between them. For each such $\delta$, there is a family of operators $K_\gamma(\delta)$, indexed by the different possible stochastic outcomes $\gamma$. The state $\ket{\Psi_{\Sigma'}}$ is then a random variable such that, for some $\gamma$
\begin{equation}
\ket{\Psi_{\Sigma'}} = \frac{K_\gamma(\delta)\ket{\Psi_\Sigma}}{\|K_\gamma(\delta)\ket{\Psi_\Sigma}\|},
\end{equation}
with probabilities determined by the norm of the resulting state. This is meant to capture the general structure of collapse dynamics without committing to a specific model such as GRW or CSL.

Second, it is noted that relativistic causal structure imposes two key conditions: evolution operators for spacelike separated regions must commute, and evolution operators associated with a region must commute with observables localized at spacelike separation from that region. These are the formal requirements meant to ensure covariance and consistency.

Third, it is argued that a viable relativistic collapse theory must have a stable vacuum. As we saw above, earlier relativistic CSL-type models tended to produce infinite energy from the vacuum. The reason, it is argued, is not merely technical: if the vacuum is Poincaré invariant, any nonzero probability of producing excitations from it would require a Poincaré-invariant probability distribution over possible excitations. But no finite invariant measure of the required kind exists over the whole mass shell. So the only Lorentz-invariant way to avoid infinite vacuum excitation is for the vacuum never to produce particles at all.

The no-go result then follows from combining vacuum stability with relativistic causality and the Reeh-Schlieder theorem. Suppose the vacuum is evolved through a bounded spacetime region $\delta$. Vacuum stability requires that all possible stochastic outcomes map the vacuum into the same ray of Hilbert space; otherwise the ensemble state would become mixed, meaning the vacuum had changed. So for almost all outcomes $\gamma,\gamma'$, there is a $c$ such that the difference
\begin{equation}
K_\gamma(\delta) - c K_\gamma'(\delta)
\end{equation}
annihilates the vacuum. But because these operators commute with all standard observables in spacelike separated regions, the Reeh-Schlieder theorem implies that if such an operator annihilates the vacuum, it annihilates every state in the standard Hilbert space. Therefore the different stochastic alternatives do not really lead to different physical states. The evolution is deterministic after all.

So the conclusion is that a stable vacuum, relativistic causality and standard degrees of freedom imply deterministic evolution. But collapse theories need genuinely indeterministic evolution to suppress macroscopic superpositions. Therefore, within Myrvold's framework, a relativistic collapse theory cannot use only standard degrees of freedom.

%%%%%%%%%%%%%%%%%%%%%%%%%%%%%%%%%%%%%%%%%%%%%%%%%%%%%%%%%%%%%%
\subsection{No-signaling theorems}
%%%%%%%%%%%%%%%%%%%%%%%%%%%%%%%%%%%%%%%%%%%%%%%%%%%%%%%%%%%%%%

The no-signaling theorem in the context of a standard quantum field theory is proven as follows. Consider such a theory defined on a globally hyperbolic spacetime $M$. To every bounded spacetime region $O \subset M$, the theory assigns an algebra $\mathcal{A}(O)$ of operators localized in $O$. Assume \emph{microcausality}:
\begin{equation}
O_1 \perp O_2 \quad \Rightarrow \quad [A,B]=0
\end{equation}
for all observables $A \in \mathcal{A}(O_1)$ and $B \in \mathcal{A}(O_2)$, where $O_1 \perp O_2$ means that every point of $O_1$ is spacelike separated from every point of $O_2$.

Now consider two spacelike separated regions $O_A \perp O_B$. Suppose Alice performs a local operation in $O_A$. Represent the nonselective operation by Kraus operators $K_\alpha$ localized in $O_A$, satisfying the completeness condition
\begin{equation}
\sum_\alpha K_\alpha^\dagger K_\alpha = I.
\end{equation}
``Nonselective'' means that we average over Alice's possible outcomes, so no conditioning on Alice's result is made.

Next, let $\Sigma$ and $\Sigma'$ be two Cauchy hypersurfaces which coincide everywhere except on the boundary of $O_A$, with $\Sigma$ to its past and $\Sigma'$ to its future (see Figure 1).
%%%%%%%%%%%%%%%%%%%%%%%%%%%%%%
\begin{figure}[t]
	\centering
	\includegraphics[height=7cm]{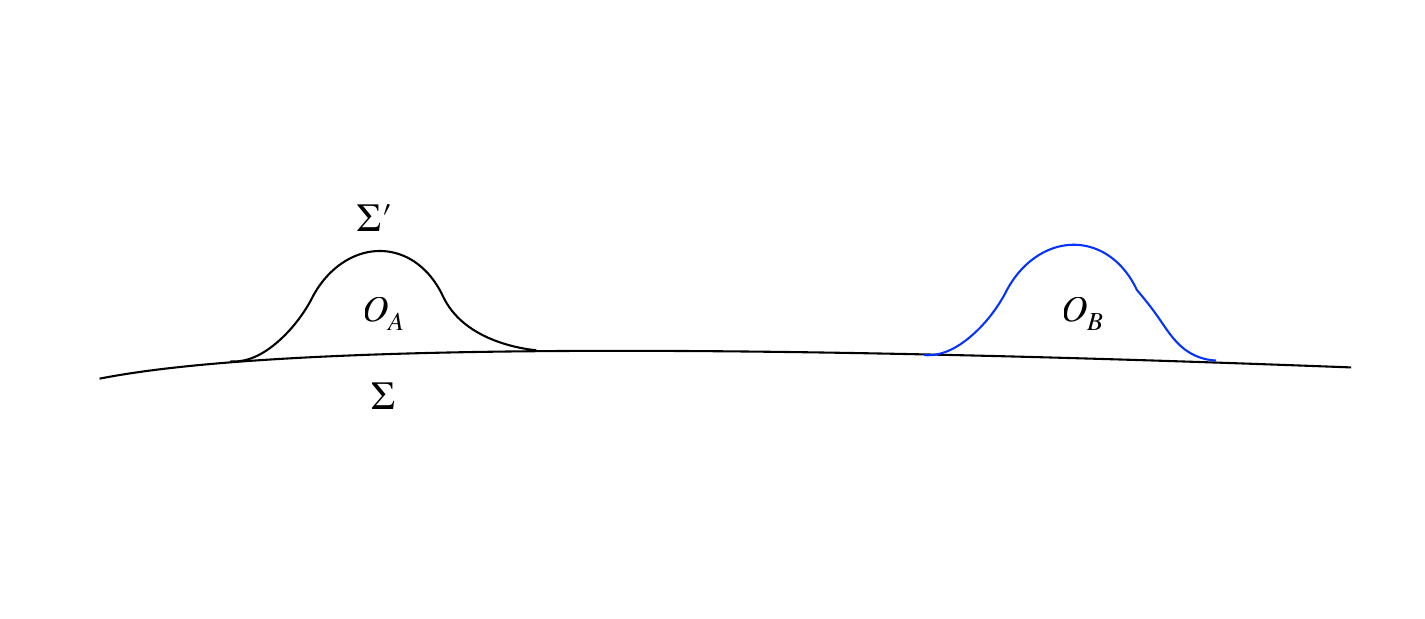} 
	\caption{The regions $O_A$ and $O_B$ are spacelike separated and the hypersurfaces $\Sigma$ and $\Sigma'$ coincide everywhere except on the boundary of $O_A$, with $\Sigma$ to its past and $\Sigma'$ to its future.}
\end{figure} \label{fig1}
%%%%%%%%%%%%%%%%%%%%%%%%%%%%%%
If the state assigned to $\Sigma$ is $\rho_\Sigma$, the state assigned to $\Sigma'$ is
\begin{equation}
\rho_{\Sigma'} = \sum_\alpha K_\alpha \rho_\Sigma K_\alpha^\dagger.
\end{equation}

Now let Bob measure an observable $E_B$ localized in $O_B$. Then, the probability of Bob's outcome after Alice's operation is
\begin{equation}
\sum_\alpha \operatorname{Tr}(K_\alpha \rho_\Sigma K_\alpha^\dagger E_B).
\end{equation}
However, using the cyclicity of the trace, microcausality between the $K_\alpha$ and $E_B$, and completeness of the $K_\alpha$, such a probability is shown to be equal to
\begin{equation}
\operatorname{Tr}(\rho_\Sigma E_B).
\end{equation}
Therefore, Alice's nonselective local operation in $O_A$ cannot change Bob's outcome probabilities in the spacelike separated region $O_B$. This is the no-signaling theorem in standard quantum field theory.

To connect this with collapse models, we note that, if, as Myrvold suggests, collapses can be represented by families of Kraus operators, then commutation between the families associated with spacelike separated regions implies no-signaling.

In \cite{Gisin1989}, Gisin offers another relevant no-signaling result. Call two different ensembles \emph{equivalent} if they are represented by the same density matrix $\rho$, and \emph{inequivalent} if they are not. Consider two initial equivalent ensembles, $E_1(0)$ and $E_2(0)$, and suppose that the proposed dynamics evolves them into inequivalent ensembles, $E_1(t)$ and $E_2(t)$.

Gisin notes that these two inequivalent ensembles can be realized remotely as follows. Suppose the two equivalent ensembles are $E_1(0)=\{x_i,|\psi_i\rangle\},E_2(0)=\{y_j,|\chi_j\rangle\}$, with
\begin{equation}
\rho= \sum_i x_i |\psi_i\rangle\langle\psi_i|= \sum_j y_j |\chi_j\rangle\langle\chi_j|.
\end{equation}
Then, one can introduce an auxiliary Hilbert space $K$ and two orthonormal sets ${|\alpha_i\rangle}$ and ${|\beta_j\rangle}$ in $K$ such that the same entangled state can be written as
\begin{equation}
|S+K\rangle= \sum_i \sqrt{x_i}|\psi_i\rangle\otimes|\alpha_i\rangle= \sum_j \sqrt{y_j}|\chi_j\rangle\otimes|\beta_j\rangle.
\end{equation}
Next, place $S$ in region $R_1$ and the auxiliary system $K$ in a distant region $R_2$. The observer in $R_2$ can choose between measuring $A$ or $B$. If $A$ is measured, then the distant system $S$ is left in state $|\psi_i\rangle$, with probability $x_i$. So, for the observer in $R_1$, the ensemble of systems $S$ is $E_1(0)$. If, instead, $B$ is measured in $R_2$, then the ensemble in $R_1$ is $E_2(0)$.

Therefore, at the initial time, these two choices are locally indistinguishable in $R_1$, since they correspond to the same density matrix $\rho$. However, if as assumed, the dynamics subsequently sends these equivalent ensembles into different density matrices $\rho_A, \rho_B$, then the observer in $R_1$ can determine which measurement was chosen in $R_2$. The distant measurement choice thereby becomes detectable superluminally. The lesson is that closed evolution at the density-matrix level is a \emph{necessary} condition for no-signaling: to avoid signaling, the future density operator must depend only on the initial density operator, not on the particular ensemble decomposition by which that density operator is realized.

Summing up, with respect to the standard no-signaling theorem, if collapses can be represented by families of Kraus operators, then a microcausality-like condition on those operators guarantees no-signaling. If, however, the collapse dynamics cannot be represented in this way, the theorem has little to say. With respect to Gisin's result, closed evolution at the level of the statistical ensemble density matrix is a necessary condition for no-signaling.

Before moving on, it is important to note that these results assume that certain observables are measurable, and in particular that they can be measured ideally in the relevant sense. In \cite{Sorkin1993} Sorkin has shown that attempts to extend the notion of ideal measurement to quantum field theory lead to a conflict with locality. This suggests that imposing some restriction on what can be measured may be an interesting strategy for circumventing these results.

%%%%%%%%%%%%%%%%%%%%%%%%%%%%%%%%%%%%%%%%%%%%%%%%%%%%%%%%%%%%%%
%%%%%%%%%%%%%%%%%%%%%%%%%%%%%%%%%%%%%%%%%%%%%%%%%%%%%%%%%%%%%%
\section{Assessing the technical challenges} 
\label{SATI}
%%%%%%%%%%%%%%%%%%%%%%%%%%%%%%%%%%%%%%%%%%%%%%%%%%%%%%%%%%%%%%
%%%%%%%%%%%%%%%%%%%%%%%%%%%%%%%%%%%%%%%%%%%%%%%%%%%%%%%%%%%%%%

In this section, we assess the technical challenges identified above. We begin with Bedingham's proposal, which shows how smearing can be introduced in a relativistic collapse model while preserving the relevant microcausality conditions, though at the cost of adding non-standard degrees of freedom. We then consider concrete methods for implementing a smearing using structures already present in relativistic spacetime, such as the Weyl tensor, the Ricci tensor, or the energy-momentum tensor. This leads to the issue of state-dependence: once smearing is tied either directly to the physical state or indirectly to geometry, the corresponding collapse operators may depend on the state. We then examine how this affects the scope of Myrvold's no-go result, before finally separating the problem of covariance from the problem of no-signaling and outlining a possible strategy for preserving covariance in discrete smeared-collapse models.

%%%%%%%%%%%%%%%%%%%%%%%%%%%%%%%%%%%%%%%%%%%%%%%%%%%%%%%%%%%%%%
\subsection{Bedingham’s proposal}
%%%%%%%%%%%%%%%%%%%%%%%%%%%%%%%%%%%%%%%%%%%%%%%%%%%%%%%%%%%%%%

We begin this section by presenting a model proposed by Bedingham in \cite{bedingham2011relativistic} that appears to meet the central technical challenge before us: constructing a framework that simultaneously satisfies bounded energy increase, covariance, and no-signaling. Bedingham’s proposed solution is to introduce an unconventional mediating field. This field has creation and annihilation operators $a^\dagger(x)$ and $a(x)$, labelled by spacetime points rather than by points on a spacelike hypersurface, satisfying
\begin{equation}
[a(x),a^\dagger(y)] = \delta^4(x-y).
\end{equation}

The matter field is coupled to this mediating field through an interaction Hamiltonian density
\begin{equation}
H_{\mathrm{int}}(x)=J(x)A(x),
\end{equation}
where $J(x)$ is a Lorentz-scalar matter-density operator and
\begin{equation}
A(x)=\int d^4y g(x,y)\bigl(a(y)+a^\dagger(y)\bigr).
\end{equation}

The collapse operator acts on a smeared number density of the mediating field,
\begin{equation}
N(x)=\int d^4y f(x,y)n(y),
\end{equation}
with
\begin{equation}
n(y)=a^\dagger(y)a(y).
\end{equation}

The physical idea is that ordinary matter first excites the mediating field, thereby leaving a smeared spacetime record of the matter distribution. Collapse then acts on this record, rather than directly on the matter field. Since the matter field and the mediating field become entangled, localization of the mediating field indirectly suppresses the corresponding branches of the matter state.

The functions $f$ and $g$ implement the smearing. Bedingham does not specify them uniquely, nor does he provide a precise general method for constructing them. Instead, he requires them to be invariantly defined and sufficiently well behaved, and suggests that the smearing functions may depend on features of the physical state itself. In this way, the smearing is not introduced by choosing an external preferred foliation or frame. Rather, the relevant local structure is picked out, at least in part, by the physical state. In the next subsection, we describe several concrete methods for implementing relativistic smearing.

The microcausality conditions are secured by imposing causal support restrictions on the smearing functions. The function $f(x,y)$, which determines what the collapse operator $N(x)$ reads, is taken to have support only when $y$ lies in the causal past of $x$. The function $g(x,y)$, which determines how matter at $x$ excites the mediating field, is taken to have support only when $y$ lies in the causal future of $x$. Thus, $N(x)$ reads from the past of $x$, while $A(x)$ writes to the future of $x$.

With these support conditions, if $x$ and $x'$ are spacelike separated, the relevant supports of $f(x,\cdot)$ and $g(x',\cdot)$ do not overlap. As a result,
\begin{equation}
[N(x),A(x')]=0,
\end{equation}
and hence
\begin{equation}
[N(x),H_{\mathrm{int}}(x')]=0
\end{equation}
for spacelike-separated points. The mediating field therefore allows the collapse interaction to be smeared without violating the commutation relations required for relativistic covariance.

Bedingham's proposal is extremely clever. By introducing a mediating field with non-standard properties, it provides a framework in which the collapse operators can be smeared without reintroducing the usual divergences, while still satisfying the relevant microcausality conditions. In this way, the model offers a genuinely covariant collapse dynamics: the evolution does not depend on a preferred foliation, since the order in which spacelike separated regions are crossed by a hypersurface does not affect the final state. The price to pay is the introduction of the non-standard pointer field.

%%%%%%%%%%%%%%%%%%%%%%%%%%%%%%%%%%%%%%%%%%%%%%%%%%%%%%%%%%%%%%
\subsection{Relativistic smearings}
%%%%%%%%%%%%%%%%%%%%%%%%%%%%%%%%%%%%%%%%%%%%%%%%%%%%%%%%%%%%%%

As we saw above, some form of smearing of the collapse operators is needed in order to avoid divergences. However, it is not straightforward to define a smearing function in a genuinely covariant way. Non-relativistic smearing prescriptions proceed by averaging the relevant operator over a small spatial region at a given time. But the distinction between spatial and temporal directions is not invariant: it depends on a choice of spacelike hypersurface. Thus, a smearing that looks natural in one frame typically amounts to selecting a preferred notion of simultaneity, thereby undermining the covariance one was trying to preserve.

One might instead try to smear over a spacetime region rather than over a purely spatial region. But in Minkowski spacetime this is also delicate. The most obvious Lorentz-invariant neighborhoods of a point are defined by the invariant interval, but such constructions end up having noncompact hyperbolic structure. This makes the smearing mathematically and physically unwieldy.

In sum, Minkowski spacetime provides very little local structure with which to define a finite, well-behaved, Lorentz-invariant smearing region. This points toward a different strategy. Rather than trying to define the smearing using only the highly symmetric structure of Minkowski spacetime, one may exploit the additional structure present in the actual spacetime we inhabit. Our spacetime is curved and contains a rich pattern of local inhomogeneities, and these features may provide covariantly defined resources with which to construct physically well-behaved smearing functions.

In what follows, we explore several possible smearing schemes that make use of covariantly defined local structure. The first is based on the algebraic classification of conformal curvature, as encoded in the Petrov classification of the Weyl tensor (see \cite{BedSud} for a similar idea). The second appeals to the algebraic classification of the Ricci tensor, usually formulated in terms of Segre (or Plebánski) types. The third relies on the physical characterization of the energy-momentum tensor, using the local matter content of spacetime to identify preferred directions. The common aim is to determine whether these structures can provide the geometric resources needed to define smearing functions without introducing a preferred foliation or any additional non-covariant background structure.

%%%%%%%%%%%%%%%%%%%%%%%%%%%%%%%
\subsubsection{Petrov classification}
%%%%%%%%%%%%%%%%%%%%%%%%%%%%%%%

The Petrov classification is a local, covariant way of classifying the algebraic structure of the Weyl tensor, \cite{StephaniEtAl2003}. At each spacetime point, the Weyl tensor determines, in general, four principal null directions. These are special null directions singled out by the curvature. When all four principal null directions are distinct, the point is algebraically generic. This is Petrov type I. In this case, the Weyl tensor has the least degenerate structure and therefore provides comparatively rich local geometric information.

By contrast, algebraically special points are those at which some of the principal null directions coincide. Such degeneracies correspond to greater symmetry, and hence to less independent geometric structure with which to distinguish local directions. The different Petrov types classify the possible ways in which this degeneracy can occur, but the important distinction for present purposes is simply between generic points, where the Weyl tensor provides maximal algebraic structure, and special points, where that structure becomes degenerate.

This distinction can be characterized invariantly. In terms of the usual complex scalar invariants $I$ and $J$ constructed from the Weyl tensor, algebraically special points are identified by the vanishing of the discriminant-like scalar
\begin{equation}
\Delta_C = I^3 - 27J^2.
\end{equation}

To use the geometric structure available in the generic case to define a covariant smearing procedure, one could proceed as follows. Given a collapse point, use the principal null directions at that point to construct canonically a unit future-directed timelike vector. Then introduce a timelike smearing length $\tau$ and, using the exponential map, choose a second point lying a proper time $\tau$ along the integral direction determined by that vector. The smearing region can then be taken to be the causal diamond with the original collapse point and this second point as its endpoints.

To construct a unit timelike vector from the geometric structure available at a point, one can proceed as follows. Suppose first that we are given three distinct future-directed null directions at a point, represented by arbitrary future-directed null vectors $k_1,k_2,k_3$. Using signature $(-+++)$, define
\begin{equation}
q_{ij}=-k_i\cdot k_j>0
\qquad
(i\neq j).
\end{equation}
We then rescale the three vectors by setting
\begin{equation}
\ell_1=\sqrt{\frac{q_{23}}{q_{12}q_{13}}} k_1,
\qquad
\ell_2=\sqrt{\frac{q_{13}}{q_{12}q_{23}}} k_2,
\qquad
\ell_3=\sqrt{\frac{q_{12}}{q_{13}q_{23}}} k_3.
\end{equation}
These rescaled vectors satisfy
\begin{equation}
-\ell_i\cdot \ell_j=1
\qquad
(i\neq j).
\end{equation}
A canonical future-directed unit timelike vector is then obtained by taking the normalized sum:
\begin{equation}
u^a=\frac{1}{\sqrt{6}}\left(\ell_1^a+\ell_2^a+\ell_3^a\right).
\end{equation}
Thus, from three distinct future-directed null directions one obtains a canonical unit future-directed timelike vector by first fixing the relative normalization of the null representatives through their pairwise inner products, and then taking the normalized sum.

Now suppose that we are given four distinct future-directed null directions, represented by arbitrary future-directed null vectors $k_1,k_2,k_3,k_4$. A practical symmetric construction is to apply the previous three-direction prescription to each of the four triples. Let
\begin{equation}
u_{(ijk)}^a
\end{equation}
denote the unit future-directed timelike vector constructed from the triple $(k_i,k_j,k_k)$ by the procedure above. One then forms their sum
\begin{equation}
U^a=
u_{(123)}^a+
u_{(124)}^a+
u_{(134)}^a+
u_{(234)}^a,
\end{equation}
and normalizes it:
\begin{equation}
u^a=\frac{U^a}{\sqrt{-U^bU_b}}.
\end{equation}
This gives a future-directed unit timelike vector constructed symmetrically from the four null directions.

Regarding applications of all this for collapse theories, the idea, is the following. Potential collapse points are distributed uniformly throughout spacetime, i.e., by a Poisson process with constant density with respect to the invariant four-volume element. At each such point, the scalar $\Delta_C$ is evaluated. If $\Delta_C=0$, no collapse occurs there. If $\Delta_C \neq 0$, a smeared collapse operator is constructed, with its smearing region defined by the procedure described above.

%%%%%%%%%%%%%%%%%%%%%%%%%%%%%%%
\subsubsection{Segre classification}
%%%%%%%%%%%%%%%%%%%%%%%%%%%%%%%

The Segre classification is a local, covariant way of classifying the algebraic structure of the Ricci tensor, \cite{StephaniEtAl2003}. Whereas the Petrov classification concerns the conformal part of the curvature, encoded in the Weyl tensor, the Segre classification concerns the part of the curvature directly tied, through Einstein's equations, to the local matter content. At each spacetime point, the Ricci tensor determines an eigenvalue problem
\begin{equation}
R^a{}_b v^b = \lambda v^a .
\end{equation}
In the algebraically generic case, $R^a{}_b$ has four distinct eigenvalues and four independent eigenvectors, that can be organized into one timelike and three spacelike eigendirections. In this situation, the Ricci tensor has the least degenerate structure and therefore provides comparatively rich local geometric information. In particular, for our purposes, in such cases the distinguished timelike direction can be used to construct a smearing region in the same way as in the Petrov case discussed above.

By contrast, algebraically special Segre types arise when this eigenstructure becomes degenerate or otherwise less informative. This can happen when eigenvalues coincide, when the tensor fails to be diagonalizable and develops nontrivial Jordan blocks, or when the eigenvectors have null rather than timelike or spacelike character. Such degeneracies correspond to greater symmetry, and hence to less independent geometric structure with which to distinguish local directions. The different Segre types classify the possible ways in which the eigenvalue and eigendirection structure of $R^a{}_b$ can degenerate, but the important distinction for present purposes is simply between generic points, where the Ricci tensor provides maximal algebraic structure, and special points, where that structure becomes degenerate.

This distinction can be characterized invariantly by the discriminant of the characteristic polynomial of $R^a{}_b$. Let
\begin{equation}
p_R(\lambda)=\det(R^a{}_b-\lambda \delta^a{}_b)
\end{equation}
be the characteristic polynomial. Then one may define
\begin{equation}
\Delta_R = \operatorname{Disc}(p_R).
\end{equation}
When the eigenvalues are written as $\lambda_1,\ldots,\lambda_4$, this discriminant is, up to convention-dependent factors,
\begin{equation}
\Delta_R=\prod_{i<j}(\lambda_i-\lambda_j)^2 .
\end{equation}
Thus $\Delta_R\neq 0$ identifies points at which the Ricci tensor has four distinct eigenvalues, while $\Delta_R=0$ identifies points at which at least two eigenvalues coincide. In this sense, $\Delta_R$ plays for the Segre classification the same role that the scalar $\Delta_C$ plays in the Petrov classification: it provides a covariant scalar test for the onset of algebraic degeneracy.

Therefore, as above, the proposal is to distribute potential collapse points uniformly throughout spacetime. At each such point, the scalar $\Delta_R$ is evaluated. If $\Delta_R=0$, no collapse occurs there. If $\Delta_R\neq 0$, a smeared collapse operator is constructed, with its smearing region defined by the procedure described in the previous section.

%%%%%%%%%%%%%%%%%%%%%%%%%%%%%%%
\subsubsection{Energy-momentum tensor}
%%%%%%%%%%%%%%%%%%%%%%%%%%%%%%%

The last option we consider, proposed in \cite{bedingham2011relativistic} (see also \cite{Durr2014}), is to use the energy-momentum tensor. More concretely, the idea is to use a Segre-type analysis on $\mathcal{T}_{ab}(x)$, defined in Eq.~\eqref{matterdensityeq} as the renormalized expectation value of the energy-momentum tensor operator, in the (pre-collapse) state assigned to the boundary of the causal past of $x$. Since such a tensor has the same tensorial structure as the Ricci tensor, the algebraic analysis given above for the Ricci tensor carries over directly. One can therefore classify $\mathcal{T}_{ab}(x)$ by its Segre type, identify the cases in which it determines a distinguished timelike direction, and then use that direction to construct a smearing region by the same procedure described above.

%%%%%%%%%%%%%%%%%%%%%%%%%%%%%%%
\subsubsection{State-dependence of the smearing prescriptions}
%%%%%%%%%%%%%%%%%%%%%%%%%%%%%%%

The smearing procedures explored above fall into two broad categories. The first uses the geometry of spacetime itself. This includes constructions based on the algebraic structure of the Weyl tensor or the Ricci tensor, where curvature is used to identify preferred local directions from which a smearing region can be defined. The second uses the physical state more directly, for instance by appealing to the energy-momentum tensor and its local algebraic structure.

At first sight, the first strategy seems to have an important advantage. Since the smearing region is constructed from spacetime geometry rather than from the quantum state, the corresponding collapse operator appears not to depend on the state itself. This is attractive because state-dependent collapse operators might threaten no-signaling, unless additional constraints are imposed. By contrast, a smearing rule based only on the background geometry seems to avoid this difficulty.

However, this distinction is less sharp than it first appears. The central lesson of general relativity is precisely that geometry is not an autonomous fixed background, but is dynamically tied to matter content. The curvature of spacetime is influenced by the distribution of energy and momentum. It is true that, if one works in quantum field theory on a fixed curved background, the quantum matter fields are treated as propagating on a spacetime geometry that is not affected by them. In that setting, a geometrically defined smearing prescription would not depend on the quantum state through gravitational back-reaction. But such a framework is understood to be an approximation. In a more complete setting, the matter content does contribute to the geometry, even if that contribution is neglected in the fixed-background treatment.\footnote{This would, of course, require some form of semiclassical framework capable of handling back-reaction (see e.g., \cite{Fully}).} Thus, even if a smearing prescription is formulated geometrically, it may still depend indirectly on the physical state of the matter fields, insofar as that state contributes to the spacetime geometry. In this sense, the first strategy does not fully escape the problem; it merely relocates the dependence from the explicit matter variables to the geometric structures they help determine.

%%%%%%%%%%%%%%%%%%%%%%%%%%%%%%%%%%%%%%%%%%%%%%%%%%%%%%%%%%%%%%
\subsection{Scope of Myrvold's no-go result}
%%%%%%%%%%%%%%%%%%%%%%%%%%%%%%%%%%%%%%%%%%%%%%%%%%%%%%%%%%%%%%

We saw that collapse models require smearing in order to avoid the divergences associated with pointlike collapse operators. In a relativistic setting, however, such smearing must be defined covariantly. The available strategies seem to fall into two broad categories: either the smearing is constructed from local geometric structure, or it is constructed from features of the physical state. The second option is explicitly state-dependent. The first may initially appear to avoid state-dependence, but in a gravitational setting this appearance is misleading. Since the local geometry is itself influenced by the matter content, a smearing prescription based on curvature will generally inherit an indirect dependence on the physical state.

Thus, in either case, the collapse operators end up depending, directly or indirectly, on the state. In Myrvold's Kraus-operator language, this means that the family of Kraus operators associated with a spacetime region is not fixed independently of the state to which it is applied. Rather, the relevant Kraus operators must be selected only after the state-dependent smearing structure has been determined. This is the sense in which relativistic regularization appears to push successful collapse models toward state-dependent collapse operators.

It is important to examine how this state-dependence affects Myrvold's result discussed above. That argument crucially depends on the dynamics being linear at the level of density-matrix evolution. Only because the evolution of density matrices is assumed to be given by a state-independent completely positive map can the demand for vacuum stability, together with the Reeh--Schlieder theorem, be used to show that all states must evolve deterministically.

If, by contrast, the family of Kraus operators associated with a spacetime region is allowed to depend on the state to which it is applied, then the key step in Myrvold's argument no longer goes through. The evolution is no longer fixed by a single state-independent map acting on arbitrary density matrices. Thus, the conclusion that vacuum stability and relativistic causality force deterministic evolution is blocked. In this sense, state-dependent collapse operators provide a possible way of evading Myrvold's no-go result.

We see, therefore, that Bedingham's proposal avoids Myrvold's conclusion not merely because it introduces non-standard fields. More fundamentally, it evades the no-go argument at an earlier stage, by employing collapse operators that are state-dependent when expressed in Kraus-operator language. Since Myrvold's result assumes a state-independent completely positive dynamics at the density-matrix level, this state-dependence prevents the argument from getting off the ground. The non-standard mediating field then plays a further role in making the smeared dynamics covariant and well behaved, but the escape from Myrvold's deterministic conclusion is already enabled by the state-dependence of the relevant Kraus operators.

There is also a second limitation of Myrvold's result: its dependence on Minkowski spacetime. A crucial premise of the argument is vacuum stability. The reason this requirement is imposed is that, in Minkowski spacetime, the vacuum is expected to be Poincaré invariant, and any stochastic production of particles from it would have to be compatible with that symmetry. Myrvold then argues that no finite Poincaré-invariant probability distribution over possible excitations exists, so the only acceptable option is for the vacuum not to produce particles at all.

But this reasoning is tied to the special structure of Minkowski spacetime. In a curved spacetime there is, in general, no global Poincaré symmetry and no uniquely defined Poincaré-invariant vacuum state. Thus the vacuum-stability requirement, at least in the form used in Myrvold's argument, cannot simply be carried over. Once that requirement is dropped or substantially reformulated, the no-go result no longer applies in the same way. This is important because the spacetime we actually inhabit is not Minkowski spacetime, but a curved spacetime. Therefore, even apart from the issue of state-dependent Kraus operators, Myrvold's theorem should be understood as a result about collapse dynamics in a highly idealized background, not as a general obstruction to relativistic collapse in realistic gravitational settings. Of course, since Minkowski spacetime is a special case of a general relativistic background, any model capable of dealing with arbitrary spacetimes must, when applied to Minkowski spacetime, satisfy the constraints imposed by Myrvold's result.

None of this is meant to diminish the importance of Myrvold's result. On the contrary, the result is valuable precisely because it identifies, with unusual clarity, a serious obstruction facing relativistic collapse theories. What the preceding discussion suggests is not that the obstruction can simply be ignored, but that its scope must be carefully understood. The theorem applies under specific assumptions: Minkowski spacetime, vacuum stability in the Poincaré-invariant sense, standard degrees of freedom, and state-independent linear evolution at the density-matrix level.

What would be extremely interesting, therefore, is either of two developments. One would be the construction of an explicit relativistic collapse model that satisfies covariance, no-signaling, bounded energy increase, and uses only standard degrees of freedom. The other would be a strengthened no-go theorem showing that such a model is impossible even when the setting is generalized to curved spacetime and even when state-dependent collapse operators are allowed. Either result would significantly clarify the status of the relativistic collapse program: the first by showing that the main technical obstacles can be overcome, the second by identifying a deeper obstruction than the one currently established.

%%%%%%%%%%%%%%%%%%%%%%%%%%%%%%%%%%%%%%%%%%%%%%%%%%%%%%%%%%%%%%
\subsection{Covariance and no-signaling}
%%%%%%%%%%%%%%%%%%%%%%%%%%%%%%%%%%%%%%%%%%%%%%%%%%%%%%%%%%%%%%

For the models presented at the beginning of Section~\ref{TC}, the microcausality conditions play a double role: they guarantee both covariance and no-signaling. If the collapse operators associated with spacelike separated regions commute, then the final state obtained by advancing a hypersurface does not depend on the order in which spacelike separated collapse regions are crossed. This secures covariance. At the same time, the same commutation relations ensure that operations performed in one region cannot be detected by measurements performed at spacelike separation. This secures no-signaling.

It is important, however, not to conflate these two requirements. Covariance and no-signaling are logically independent constraints. Covariance requires the absence of non-dynamical background structure in the formulation of the dynamics. No-signaling, by contrast, concerns the operational impossibility of using the dynamics to transmit controllable information between spacelike separated regions. There is no general reason why satisfying one should automatically imply satisfying the other.

In particular, as Maudlin emphasized in \cite{maudlin2011quantum}, nonlocality, and even superluminal signaling, are not by themselves incompatible with covariance. A theory may allow correlations or influences between spacelike separated events without selecting a preferred frame, provided that the laws governing those correlations are formulated in a covariant way. The usual worry is not that signaling immediately violates covariance, but that signaling may lead to causal inconsistencies or closed causal loops. Whether that occurs depends on the detailed structure of the theory, not merely on the presence of nonlocal dependence.

In this section, we therefore separate the problem of covariance from the problem of no-signaling. Our aim is to explore alternative ways of satisfying covariance, even if those strategies do not by themselves automatically guarantee no-signaling. This is not necessarily to abandon the no-signaling requirement, but to clarify the structure of the technical problem. One may first ask whether a collapse dynamics can be formulated without a preferred foliation or non-covariant background structure, and only then ask whether the resulting dynamics also prevents operational signaling.

Consider, in particular, discrete collapse models. In such models, the relevant collapse events form a discrete set in spacetime. The usual way to secure covariance is to impose a microcausality condition: collapse operators associated with spacelike separated regions must commute, so that the final state does not depend on the order in which those regions are crossed by a spacelike hypersurface. As we saw above, however, the need to introduce smearing makes this condition difficult to satisfy. Once collapse operators are smeared over spacetime regions, ensuring that all operators associated with spacelike separated collapse events commute becomes a nontrivial constraint.

An alternative strategy is to relax this demand. Instead of requiring that all spacelike separated collapse operators commute, one could try to define a canonical order for the relevant collapse events. If this order is determined covariantly by the spacetime structure itself, then the dynamics need not depend on an arbitrary choice of foliation. The state associated with a hypersurface would be obtained by applying the collapse operators in the canonical order, rather than in the order induced by a chosen slicing of spacetime.

More cautiously, one need not require a canonical order for the entire set of collapse events. It may be enough to define such an order only for those events whose relative ordering would otherwise threaten the covariance of the dynamics, namely spacelike separated collapse events associated with noncommuting operators. If two collapse operators commute, their order is irrelevant. The ordering problem arises only when their noncommutativity makes the final state depend on which collapse is applied first. In those cases, a covariantly defined canonical ordering could replace microcausality as the mechanism securing covariance.

The proposal is the following. Potential collapse points are distributed uniformly throughout spacetime, and one of the smearing prescriptions described above is used to assign a smearing region to each of them. If two collapse points are timelike related, their order is fixed by the causal structure. If two collapse points are spacelike separated and their corresponding smearing regions are also completely spacelike separated, then the associated collapse operators commute, so their order is irrelevant.

The relevant cases, therefore, are those in which the collapse points themselves are spacelike separated, but their associated smearing regions are not. Among these, there are cases in which one smearing region contains points in the causal future of the other, but no points in its causal past (see Figure 2).
%%%%%%%%%%%%%%%%%%%%%%%%%%%%%%
\begin{figure}[t]
	\centering
	\includegraphics[height=7cm]{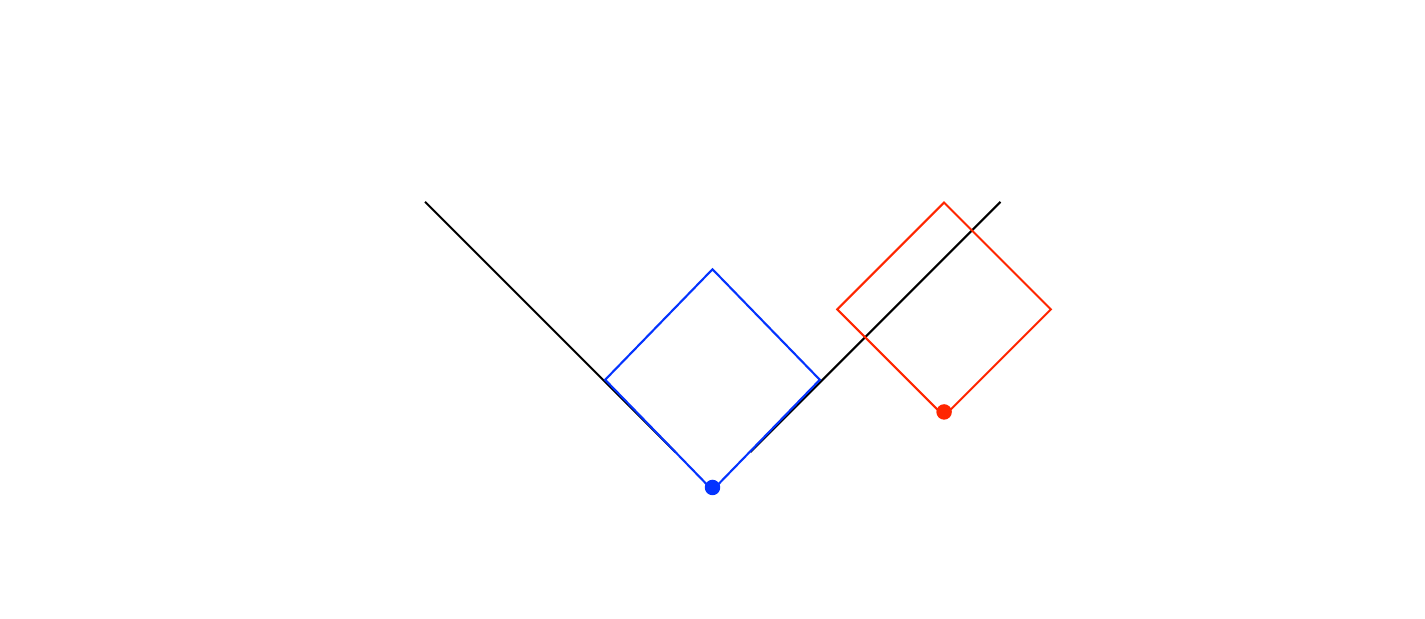} 
	\caption{The collapse points are spacelike separated and one smearing region contains points in the causal future of the other, but no points in its causal past.}
\end{figure} \label{fig2}
%%%%%%%%%%%%%%%%%%%%%%%%%%%%%%
In such cases, there is a natural ordering. If, for instance, the smearing region $S_y$ contains points to the future of $S_x$, but no points to its past, then the collapse associated with $y$ is taken to occur after the collapse associated with $x$.

The only remaining case is one in which the two smearing regions are mutually related in causal order: each contains points both to the future and to the past of the other. This situation arises, in particular, when the two smearing regions overlap (see Figure 3).
%%%%%%%%%%%%%%%%%%%%%%%%%%%%%%
\begin{figure}[t]
	\centering
	\includegraphics[height=7cm]{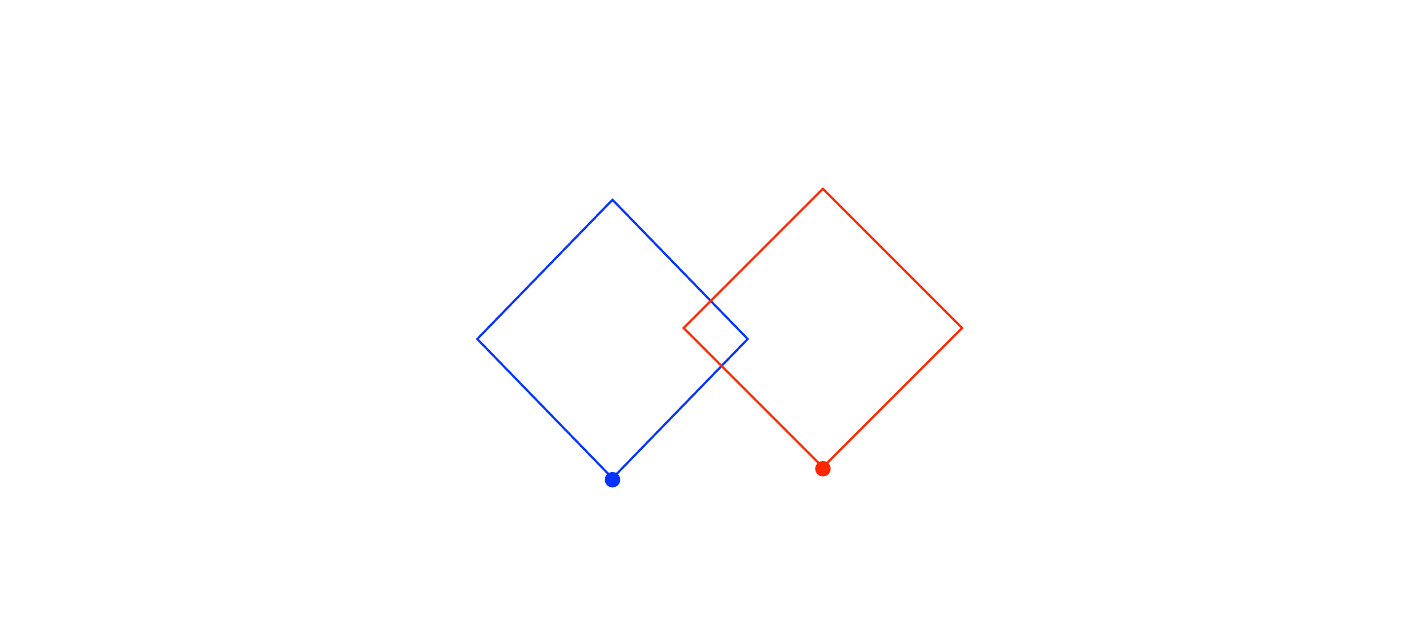} 
	\caption{The smearing regions are mutually related in causal order: each contains points both to the future and to the past of the other.}
\end{figure} \label{fig3}
%%%%%%%%%%%%%%%%%%%%%%%%%%%%%%
In such cases, the proposal is simply to eliminate the associated collapses. That is, collapse events whose smearing regions do not admit a covariantly defined ordering are discarded rather than forced into an arbitrary foliation-dependent order.

One might worry that this procedure could eliminate so many candidate events that no collapses occur at all. Whether this happens, however, must be assessed case by case, since it depends on the density of potential collapse points and on the detailed geometry of the associated smearing regions. At least in situations in which the smearing regions exhibit sufficient alignment, the problem should not arise: many pairs of regions will either be completely spacelike separated, so that their order is irrelevant, or will admit a natural causal ordering. In such cases, it seems that only very few collapses would be discarded.

In sum, in discrete collapse models with smeared operators, covariance can be preserved by assigning collapse events a covariantly defined order whenever their smearing regions make the order relevant, and by discarding the exceptional cases in which no such order is available. This strategy separates the problem of covariance from the problem of no-signaling. It offers a possible route to covariant dynamics even when smearing makes strict microcausality difficult to maintain, while leaving the signaling issue as a further, independent question.

%%%%%%%%%%%%%%%%%%%%%%%%%%%%%%%%%%%%%%%%%%%%%%%%%%%%%%%%%%%%%%
%%%%%%%%%%%%%%%%%%%%%%%%%%%%%%%%%%%%%%%%%%%%%%%%%%%%%%%%%%%%%%
\section{Conclusions} 
\label{Con}
%%%%%%%%%%%%%%%%%%%%%%%%%%%%%%%%%%%%%%%%%%%%%%%%%%%%%%%%%%%%%%
%%%%%%%%%%%%%%%%%%%%%%%%%%%%%%%%%%%%%%%%%%%%%%%%%%%%%%%%%%%%%%

The aim of this paper has been to reassess the widespread thought that objective collapse theories are fundamentally incompatible with relativity. We have argued that this conclusion is not warranted. The main conceptual objections identify real issues, but they do not establish incoherence. Collapse need not require a preferred foliation if the dynamics is formulated in terms of states assigned to arbitrary Cauchy hypersurfaces. The hypersurface-dependence of those states need not imply frame-dependent local facts, provided that physical properties are read from covariantly defined local beables. Nor does the failure of narratability, or the presence of nonlocal dependence, by itself show that relativistic collapse is conceptually impossible.

The central problems are technical. A viable relativistic collapse theory must combine bounded energy increase, covariance, consistency, and no-signaling. Pointlike collapse operators can satisfy the relevant microcausality conditions, but they generate divergent energy production. Smearing avoids this divergence, but it makes covariance and no-signaling much harder to secure. Bedingham's model addresses this problem by introducing a non-standard mediating field, allowing the collapse dynamics to be smeared while preserving the required commutation relations. This is an important achievement, though it comes at the cost of adding non-standard degrees of freedom.

We also considered concrete methods for implementing a covariant smearing using only structures already present in the relativistic setting. The Petrov classification of the Weyl tensor, the Segre classification of the Ricci tensor, and the algebraic structure of the energy-momentum tensor may provide local, covariantly defined resources for constructing finite smearing regions. However, these strategies also point toward state-dependence. Smearing prescriptions either depend explicitly on the physical state or depend on spacetime geometry, which in a gravitational setting is itself tied to matter content. Thus, in Kraus-operator language, relativistic regularization appears to push collapse models toward state-dependent collapse operators.

With this, we have clarified the scope of Myrvold's no-go result. The theorem identifies a serious obstruction for relativistic collapse models constrained to Minkowski spacetime, assuming vacuum stability, standard degrees of freedom, and state-independent linear evolution at the density-matrix level. But if the relevant Kraus operators depend on the state, or if the setting is generalized to curved spacetime, the argument no longer applies in the same way. This does not diminish the result; it shows where further work is needed. What remains open is either to construct a model satisfying covariance, no-signaling, bounded energy increase, and standard degrees of freedom, or to prove a stronger no-go theorem ruling out such models even with state-dependent collapse operators in curved spacetime.

Finally, we emphasized that covariance and no-signaling are conceptually independent. In simple pointlike models, the same microcausality conditions secure both. Once collapse operators are smeared, this connection may fail. We suggested that covariance might still be preserved in discrete collapse models by assigning a covariantly defined order to noncommuting collapse events, while treating no-signaling as a further independent constraint.

Relativistic collapse theories therefore remain technically viable, but their prospects depend on whether the remaining obstacles can be handled within a precise dynamics. Existing proposals suggest possible ways of combining bounded energy increase, covariance, and an acceptable account of signaling, though it remains unclear how far these strategies can be pushed. The central task is to determine whether these requirements can be satisfied simultaneously, either by models of the sort already available or by suitable refinements of them.\footnote{During the last stages of the elaboration of this manuscript we were made aware of the work \cite{Gundhi} which considers an approach towards the construction a relativistic collapse theory framed in a quantum field theoretical setting that makes use of non-Markovian noise, together with normal order prescription, in order to avoid infinite energy production. Unfortunately, it is unclear whether their master equation can be derived from a consistent and relativistic state evolution equation.}

%%%%%%%%%%%%%%%%%%%%%%%%%%%%%%%%%%%%%%%%%%%%%%%%%%%%%%%%%%%%%%
%%%%%%%%%%%%%%%%%%%%%%%%%%%%%%%%%%%%%%%%%%%%%%%%%%%%%%%%%%%%%%
%\section*{Acknowledgments} 
%%%%%%%%%%%%%%%%%%%%%%%%%%%%%%%%%%%%%%%%%%%%%%%%%%%%%%%%%%%%%%
%%%%%%%%%%%%%%%%%%%%%%%%%%%%%%%%%%%%%%%%%%%%%%%%%%%%%%%%%%%%%%

%%%%%%%%%%%%%%%%%%%%%%%%%%%%%%%%%%%%%%%%%%%%%%%%%%%%%%%%%%%%%%
%%%%%%%%%%%%%%%%%%%%%%%%%%%%%%%%%%%%%%%%%%%%%%%%%%%%%%%%%%%%%%
\bibliographystyle{plain}
\bibliography{ref.bib}
%%%%%%%%%%%%%%%%%%%%%%%%%%%%%%%%%%%%%%%%%%%%%%%%%%%%%%%%%%%%%%
%%%%%%%%%%%%%%%%%%%%%%%%%%%%%%%%%%%%%%%%%%%%%%%%%%%%%%%%%%%%%%	
\end{document}